\documentclass{aa}
\usepackage{epsfig}
\usepackage{graphicx}
\usepackage{mathtools}
\usepackage{txfonts}
\usepackage{lscape}
\usepackage{hyperref}
\usepackage{color}

\bibpunct{(}{)}{;}{a}{}{,}

\begin{document}
\title{
Three-dimensional dust mapping in the Orion Complex, combining Gaia-TGAS, 2MASS, and WISE
}
\subtitle{}

\author{S. Rezaei Kh.\inst{\ref{inst1}} \and C.A.L. Bailer-Jones\inst{\ref{inst1}} \and E. F. Schlafly\inst{\ref{inst2}} \and M. Fouesneau\inst{\ref{inst1}}\\
}
\institute{Max Plank Institute for Astronomy (MPIA),  K\"onigstuhl 17, 69117 Heidelberg, Germany \label{inst1} \and Lawrence Berkeley National Laboratory, One Cyclotron Road, Berkeley, CA 94720, USA \label{inst2}}

\def\teff{$T_{\rm eff}$}

\def\expec{{\mathrm E}}
\def\cov{{\mathrm Cov}}

\def\trans{^\mathsf{T}}
\def\inv{^{-1}}
\def\mby{\!\times\!}
\def\mplus{\!+\!}

\def\geonj{g_{n,j}}
\def\gvec{{\mathbf g}}
\def\geomat{{\mathrm G}}

\def\likecovN{{\mathrm V}_N}

\def\extn{a_n}
\def\extsdn{\sigma_n}
\def\extvecN{{\mathbf a}_N}
\def\extcovN{\Sigma_N}
\def\estextn{f_n}
\def\estextvecN{{\mathbf f}\!_N}

\def\rhonj{\rho_{n,j}}
\def\rhovecJ{{\boldsymbol \rho}_J}
\def\rhovecJp{{\boldsymbol \rho}_{J+1}}
\def\rhoJp{\rho_{J+1}}

\def\gpcov{{\mathrm C}}
\def\gpcovJ{{\mathrm C}_J}
\def\gpcovJp{{\mathrm C}_{J+1}}
\def\gpcovel{c}
\def\kvecJ{{\mathbf k}_J}
\def\kJp{k}

\def\mmatJ{{\mathrm M}_J}
\def\mvecJ{{\mathbf m}_J}

\def\bvecJ{{\mathbf b}_J}
\def\hvecJ{{\mathbf h}_J}
\def\rmatJ{{\mathrm R}_J}

\def\rvec{{\mathbf r}}
\def\ofo{{\mathcal O}}


\abstract{We present a map of the three-dimensional (3D) distribution of dust in the Orion complex. Orion is the closest site of high-mass star formation, making it an excellent laboratory for studying the interstellar medium and star formation. We use data from the Gaia-TGAS catalogue combined with photometry from 2MASS and WISE to get the distances and extinctions of individual stars in the vicinity of the Orion complex. We adopt the non-parametric method of \citet{Rezaei_Kh_17} to infer the probability distribution function of the dust densities at arbitrary points throughout the region. We map the dust distribution towards different parts of the Orion complex 
and find the distance and depth of the cloud compatible with other recent works which show the ability of the method to be applicable on the local molecular clouds to map their 3D dust distribution. We also demonstrate the danger of only using colours of stars to derive their extinctions without considering further physical constraints like the colour-magnitude diagram (CMD).}


\maketitle

\section{Introduction}

In the interstellar medium (ISM), the dust-to-gas ratio is only one percent, and gas itself is just a few percent of the total mass of the Galaxy. Dust, therefore, is only a tiny fraction of matter in the Galaxy. Yet it plays an important role. It absorbs light at short wavelengths, scattering and re-emitting it at long wavelengths, thereby playing a major role in shaping the interstellar radiation field. Dust shields molecular hydrogen from destruction, and catalyses its formation, thus allowing stars to form. Studying the properties of dust, such as its three-dimensional (3D) distribution, is crucial to understanding the formation and evolution of galaxies.

The Orion molecular complex is one of the most studied regions in the Galaxy. Containing M42, it is the closest H II region and closest site of massive star formation \citep{ODell01}. 
Estimates of the distance to the Orion nebula cluster, the active star-forming regions in Orion which extends over 200 pc both in radial direction and in the plane of the sky \citep[e.g.][]{Brown94, Bally08}, range from 347 pc to 483 pc 
using optical and near infrared photometry and colour-magnitude diagrams \citep{Jeffries07}. Other methods have found different distances to the cloud of 440 pc and 392 pc \citep{Jeffries07}, 480 pc \citep{Genzel81}, 434 pc and 387 pc \citep{Kraus07} and 414 pc \citep{Menten07}. Recently, \cite{schlafly15} mapped the dust in Orion using Pan-STARRS1 photometry, revealing a 150 pc depth dust ring extending from around 415 pc to 550 pc in distance from the Sun.

Different techniques have been developed for mapping dust distributions in 3D \citep[e.g.][]{Vergely10, SaleM14, Hanson14, Lallement14, Green15, Hanson16, Rezaei_Kh_17, Capitanio17}. These works are all closely related in that they use estimates of the reddening and distance to stars in the Milky Way to infer the three dimensional distribution of dust.  They distinguish themselves from one another in the sources of data adopted, the techniques for translating those data into reddening and distance estimates, and the approaches used to infer the 3D dust distribution using those estimates. In an upcoming paper (Rezaei Kh et al. in prep.) we discuss the developments and improvements of our model. Here we report on the use of this model to map the dust towards Orion.

The paper is organised as following: in section \ref{data} we describe our data sets and how we derive the stellar extinctions and distances. In section \ref{method} we give an overview of the method and the parameters used to set up the model. we present our inferred dust maps in section \ref{Orion} then discuss them in section \ref{discussion} where we mention how they are affected by the inclusion of other data, and how the analysis can be extended in the future. 

\section{Data}\label{data}

As explained in \cite{Rezaei_Kh_17}, we need to have the 3D positions of the stars, i.e.\ their longitudes, latitudes, and distances (l, b, d), together with their l.o.s extinction measurements to derive dust densities. Positions and parallaxes of around two million stars were published as the 
Tycho-Gaia astrometric solution \citep[TGAS;][]{Michalik15,Lindegren16} 
as part of the first Gaia data release \cite[Gaia DR1;][]{Gaia_collaboration_a}.
\cite{Astra16} used this to infer distances, and we use their published distances (those inferred with the Milky Way prior without the additional systematic errors) together with the Gaia positions to determine the 3D positions of stars.

We then estimate extinctions for these stars using the Rayleigh-Jeans Colour Excess method (RJCE) \citep{Majewski11}, which uses near and mid infrared (NIR and MIR respectively) data to estimate the l.o.s $K_{s}$-band extinction. Since the dereddened colour (H - [4.5${\mu}$])$_{\circ}$ is almost constant for different types of stars, extinction values can be calculated as
\begin{equation}
{A}_{{k}_{s}} = 0.918 (H - [4.5\mu] - 0.08)
\label{eqn:dustint}
\end{equation}
with an approximate extinction uncertainty of less than 0.1 mag, where $H - [4.5\mu]$ is the measured colour \citep{Majewski11}.
We use NIR photometry (J, H, and $K_{s}$) from the Two Micron All-Sky Survey \citep[2MASS,][]{Skrutskie06} and the MIR photometry (W1 and W2) from WISE \citep{Wright10}. Both catalogues are cross-matched with TGAS in the Gaia archive\footnote{https://gea.esac.esa.int/archive/}, making estimation of the extinctions for our sample rather trivial.

Since we do not currently take into account distance uncertainty in our model, we only select stars with fractional distance uncertainties below 0.15. This gives around 650\,000 stars out to 700 pc.
\begin{figure} 
\begin{center}
\includegraphics[width=0.50\textwidth, angle=0]{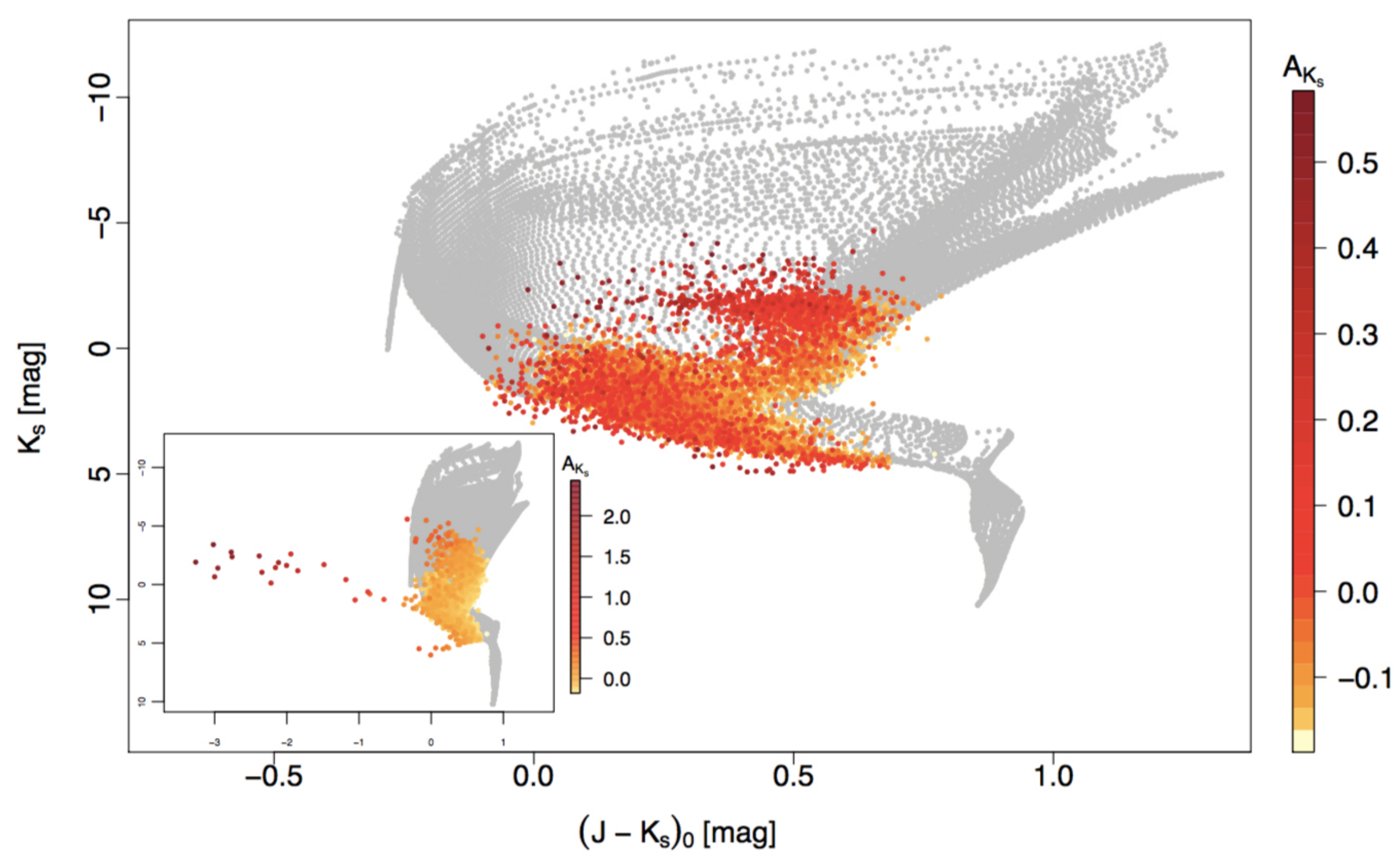}
\caption{Colour-magnitude diagram (absolute $K_{s}$ band magnitude vs. dereddened colour $(J - K_{s})_0$) for the theoretical models (grey points) and our data (in colour). The colour indicates the derived extinctions from the RJCE method. We cut some stars which are not compatible with the theoretical models, as can be seen in the inset plot (which is the same, but extending to bluer colours). Note that the colour scales are different in two plots and the rare stars have spuriously high extinctions of up to about 2 mag in the $K_{s}$ band.}
\label{fig:CMD}
\end{center}
\end{figure}
From this sample, we select stars towards the Orion complex, defined here as the region $187.5^{\circ} < l < 218^{\circ}$ and $-25^{\circ} < b < -4^{\circ}$. This leaves around 12\,000 stars within distance of 100--650 pc which we use as the input data. To check the derived extinctions we can look at the distribution of stars in the colour-magnitude diagram (CMD). Figure \ref{fig:CMD} shows the absolute $K_{s}$ band magnitude vs. dereddened colour $(J - K_{s})_0$ for the input data with RJCE extinction values (coloured points) on top of the theoretical isochrones with no extinction (grey points), for which we have used PARSEC 1.2S\footnote{http://stev.oapd.inaf.it/cgi-bin/cmd} \citep{Tang14, Chen15} to compute 2MASS JH$K_{s}$ dereddened photometry for solar metallicity stars \citep[$Z_\odot = 0.0152$,][]{Bressan12}, using the extinction law of \citet{Cardelli89} and \citet{ODonnell94} with $R_{V}$=3.1 \citep{Girardi08}.
As can be seen in the sub-plot at the lower left corner of figure \ref{fig:CMD}, a small number of points have spuriously large $A_{{K}_{s}}$ and accordingly extremely blue inferred intrinsic $J - K_{s}$ colours. Examining these stars in detail reveals that almost all of them are variable stars and young stellar objects.  We choose to eliminate them by discarding stars with $J - K_{s}$ < - 0.3 mag (this removes 19 stars). We discuss these stars and their effects on the predictions in section \ref{high}.
\begin{figure} 
\begin{center}
\includegraphics[width=0.50\textwidth, angle=0]{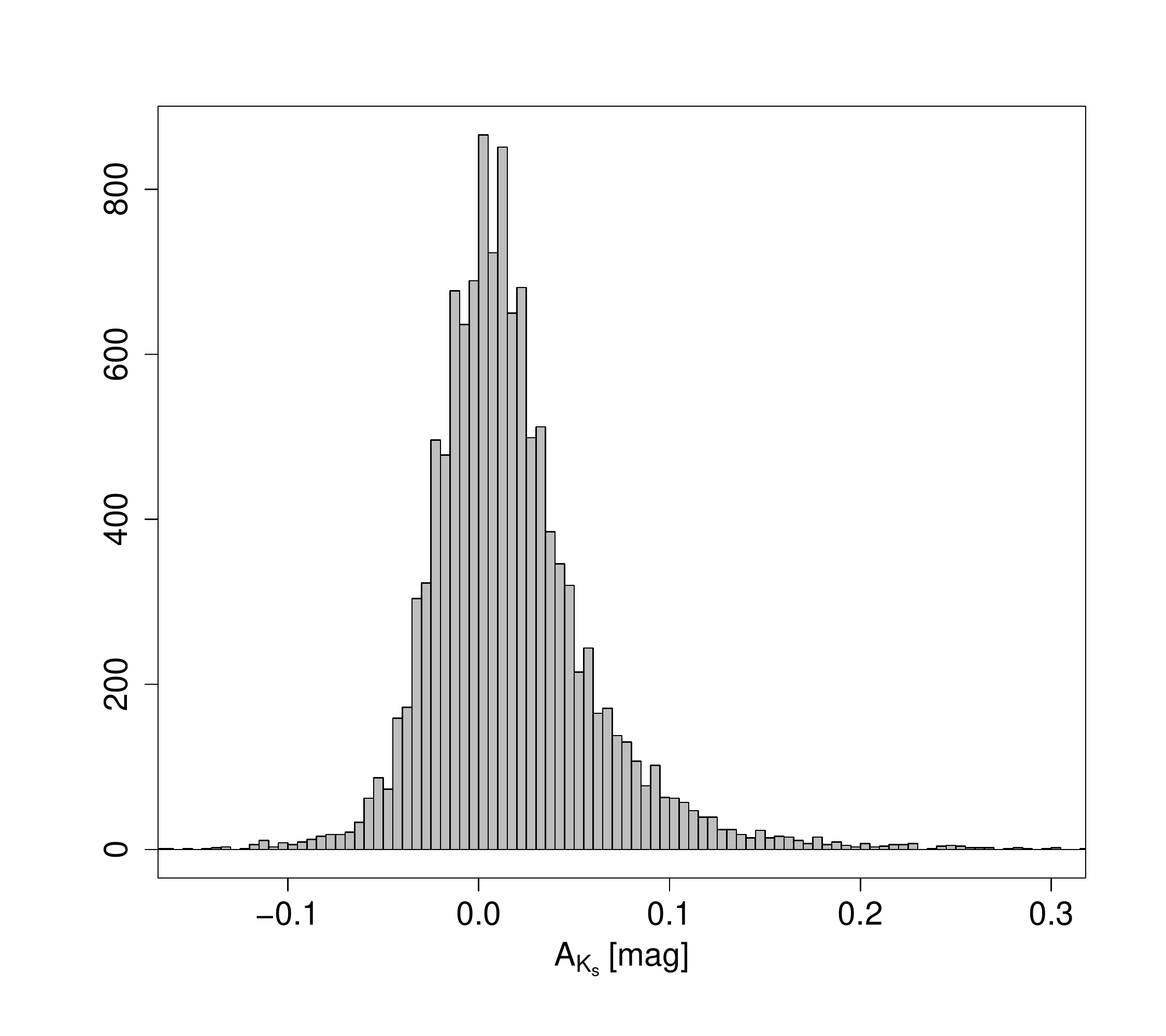}
\caption{Histogram of extinction of stars. For illustrative purposes, the main plot is limited to the range -0.2 to 0.3 mag while very few stars have extinctions up to about 0.6 mag.}
\label{fig:hist}
\end{center}
\end{figure}

As the measured extinctions are noisy, they can be negative, and in fact about $40\%$ of our sample have negative extinctions. Figure \ref{fig:hist} shows the histogram of extinction for all input stars. The $A_{{K}_{s}}$ values go up to about 0.5 magnitudes but most of the stars have extinction values lower than about 0.2 magnitudes (around 2 magnitudes in V band). Unlike in our previous work \citep{Rezaei_Kh_17}, we use all these data (including negative extinctions) in our model to predict the underlying dust density. Subsequently, in section \ref{negative}, we discuss the effects of input stars with negative extinctions on the predictions by running the model only for positive extinctions. The RJCE method assumes three sources of uncertainty in the predicted extinctions: intrinsic scatter in the stellar colour, photometric uncertainty and the uncertainties in the extinction law that they used. Combining all these sources, they estimate extinction uncertainty of less than 0.11 mag in $K_{s}$ band for a typical individual star, while being more precise for red clump stars and giants \citep{Majewski11}. Since we do not have any selection on types of stars, we assume the upper limit of the extinction uncertainty for all stars in our sample. We discuss the effects of different extinction uncertainties on our predictions in section \ref{error}.

\section{Method}\label{method}
\begin{figure*}
\resizebox{\hsize}{!}{\includegraphics[clip=true]{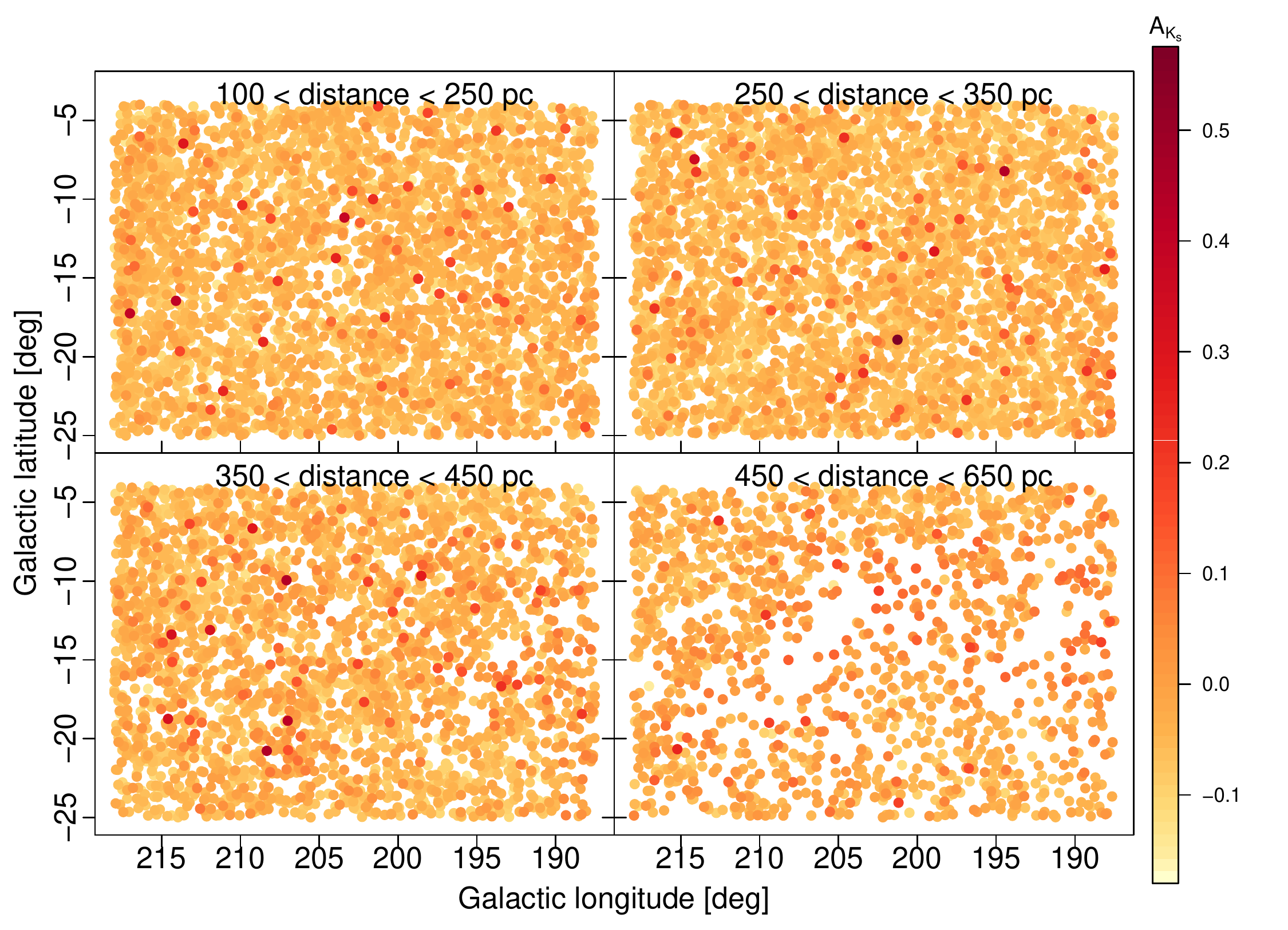}}
\caption{Galactic latitude vs. longitude of input stars towards Orion complex ($187.5^{\circ} < l < 218^{\circ}$ and $-25^{\circ} < b < -4^{\circ}$) for four different distance slices. Colour shows the extinctions. Most of stars have extinction values lower than about 0.3 magnitudes but there are some higher extinction values up to about 0.6 magnitudes. The Orion cloud can be identified in the lower right panel (high extinction stars surrounding a slight void), but it is hard to see any trend in the other three panels.
}
\label{fig:input_orion}
\end{figure*}

Our method has been described in some detail in 
\cite{Rezaei_Kh_17}. It takes line-of-sight (l.o.s) extinction measurements and distances towards stars at different positions in 3D space, then combines them using a Gaussian process model to allow us to infer the most probable 3D distribution of the dust in the Galaxy at any specified point.

As explained in \cite{Rezaei_Kh_17}, we divide each l.o.s into a set of contiguous 1D cells and assume a uniform dust density within each cell. The extinction towards each star is then approximated as the sum of the dust densities multiplied by their corresponding cell lengths, while the extinction uncertainties are taken into account. In this work, our cells are 20 pc long 
which is of the order of the typical star separations in the input data. These cells along different lines-of-sight are then connected in 3D space using a Gaussian process prior. 

This model comprises two hyperparameters, ${\lambda}$ and ${\theta}$, which determine the correlation scale and amplitude of the dust density variations, respectively. Here we adopt $\lambda=100$\, pc (a few times the typical separation of the stars) and $\theta=2.6 {\times} {10}^{-8}$ ${pc}^{-2}$, which is calculated based on the variance in the input density distribution \citep[see equation 16 in][]{Rezaei_Kh_17} in both case of including and excluding negative extinctions. We also allow a non-zero mean for our Gaussian process prior, and set it to the average dust density according to the input data. As our dust density is just extinction per unit distance, we compute the extinction divided by distance for each star, then take the mean of these as the global mean for the Gaussian process.
The non-zero mean has the advantage of representing the same properties of the ISM as the input data suggests rather than the zero-mean. The corresponding mean density for the Orion region is 0.7 ${\times} {10}^{-4}$ $pc^{-1}$ when including all data; however, by removing negative extinctions from the sample, we increase this value to 1.6 ${\times} {10}^{-4}$ $pc^{-1}$. Here we use all data including negative extinctions and will discuss the results using only positive extinctions in section \ref{negative}.

Having set these hyperparameters, we can predict the probability distribution function of the dust densities for any arbitrary point in the 3D space.

\section{Three-dimensional structure in Orion}\label{Orion}

\begin{figure*}
\resizebox{\hsize}{!}{\includegraphics[clip=true]{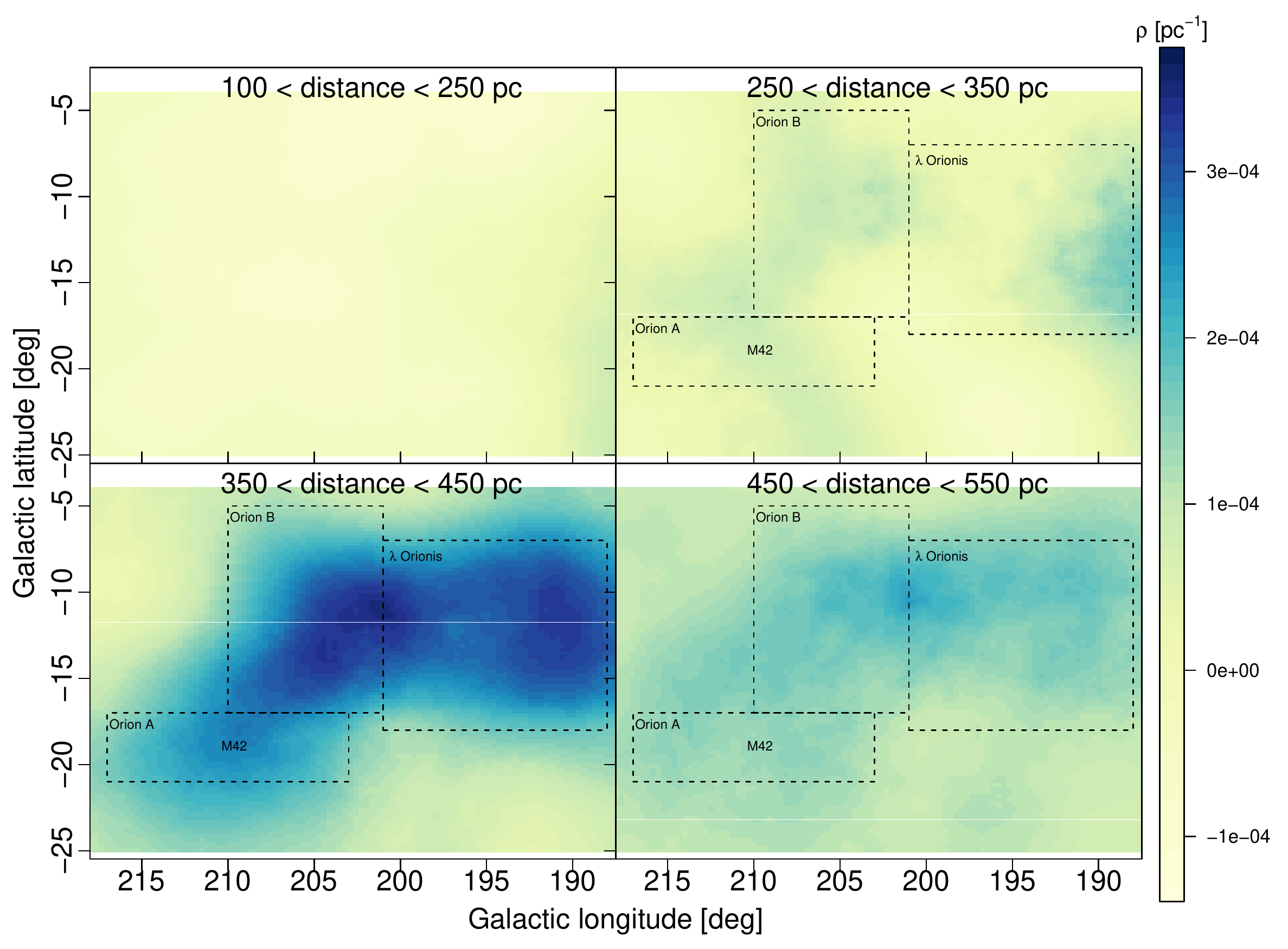}}
\caption{Dust density predictions for 50\,000 points randomly distributed towards the Orion complex (as in figure \ref{fig:input_orion}) where for plotting we take the mean of each $0.3^{\circ} {\times} 0.3^{\circ}$ grid, then smooth the plot with a Gaussian function with a scale parameter of 0.4 to cover ``empty'' pixels. Each panel shows a different distance range. Green and blue colours show the points with higher dust densities. Orion A, Orion B, and $\lambda$ Orionis are marked by dashed lines \cite{Lombardi11}.}
\label{fig:pred_2by2_orion}
\end{figure*}

Figure \ref{fig:input_orion} shows the spatial distribution of the input data (latitude vs.\ longitude) for four different distance slices (less that 250 pc, then 100-pc slices up to 450 pc and 450--650 pc slice), colour-coded by their extinction values. We use these values as the input data for our model, assuming an extinction uncertainty of 0.1 mag for all stars (see section \ref{data}). Using this we calculate the underlying dust densities for 50\,000 points randomly distributed in space towards the Orion complex with distances between 100 pc and 600 pc.

\begin{figure*}
\resizebox{\hsize}{!}{\includegraphics[clip=true]{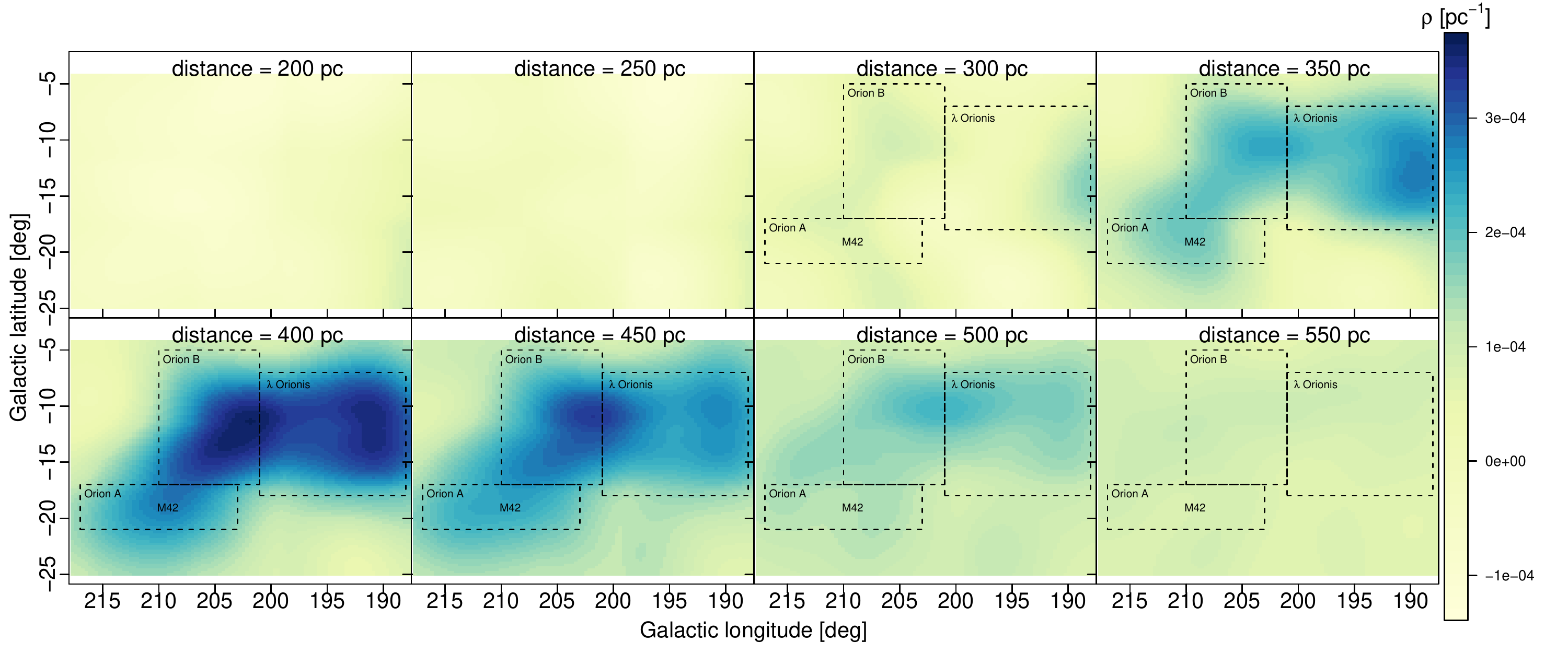}}
\caption{Dust density predictions for 10\,000 points at fixed distance distributed on a regular grid. There is no smoothness or interpolation for plotting and the features are the direct outcome of our predictions. Orion A, Orion B, and $\lambda$ Orionis are marked by dashed lines \cite{Lombardi11}.}
\label{fig:pred_2by4_orion}
\end{figure*}

Figure \ref{fig:pred_2by2_orion} shows the dust density predictions for the same panels as figure \ref{fig:input_orion}. Each panel represents the mean of the dust density predictions for the corresponding distance range. There is not much dust seen out to 250 pc (top left panel). The dust density starts to increase from the second (250--350 pc) up to the last (450 -- 550 pc) distance range, suggesting that the Orion cloud in some parts is extended for about 200--300 pc. In the lower left panel, the clear detection of Orion A and Orion B towards $\lambda$ Orionis is evident \citep[as marked by dashed lines from][]{Lombardi11}, where Orion B and $\lambda$ Orionis seem to extend to distances of larger than 450 pc. Figure \ref{fig:pred_2by4_orion}, on the other hand, shows the exact value of the predictions at fixed distances (in 50 pc steps). The first traces of over-densities starts appearing at 300 pc and extends up to at least 500 pc. Note that we restrict our data to the fractional distance uncertainty of < 0.15 which in average is about 50 pc at 400 pc distance.

\begin{figure*}
\resizebox{\hsize}{!}{\includegraphics[clip=true]{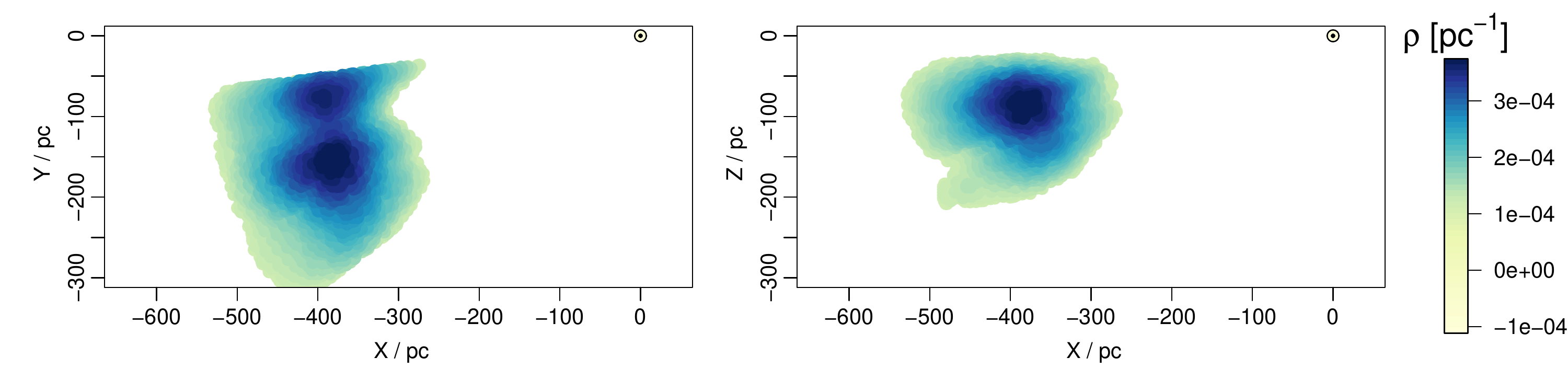}}
\caption{Two Cartesian projections of the 3D dust distributions in Orion. Only dust densities higher than the mean density of the predictions (1.2 ${\times} {10}^{-4}$ pc$^{-1}$) are shown. The Sun is at (X, Y, Z) = (0 , 0, 0), with X increasing towards the Galactic centre and Z point to the North Galactic pole, perpendicular to the Galactic disk.
}
\label{fig:3D_orion}
\end{figure*}

In order to see the structures and the locations of different parts, we need to look at their spatial distributions. Figure \ref{fig:3D_orion} shows 2D projections of the dust from two directions: top-down view (i.e.\ looking from above the Galactic plane) and side view. For better visualisation, the plot only shows points with dust predictions higher than the mean density of the predictions (1.2 ${\times} {10}^{-4}$ pc$^{-1}$). 
We see that the front part of the higher density region starts as close as 300 pc, and extends to 550 pc at the direction of the Orion A, Orion B and $\lambda$ Orionis. 

\begin{figure} 
\begin{center}
\hspace*{-2em}\includegraphics[width=0.50\textwidth, angle=0]{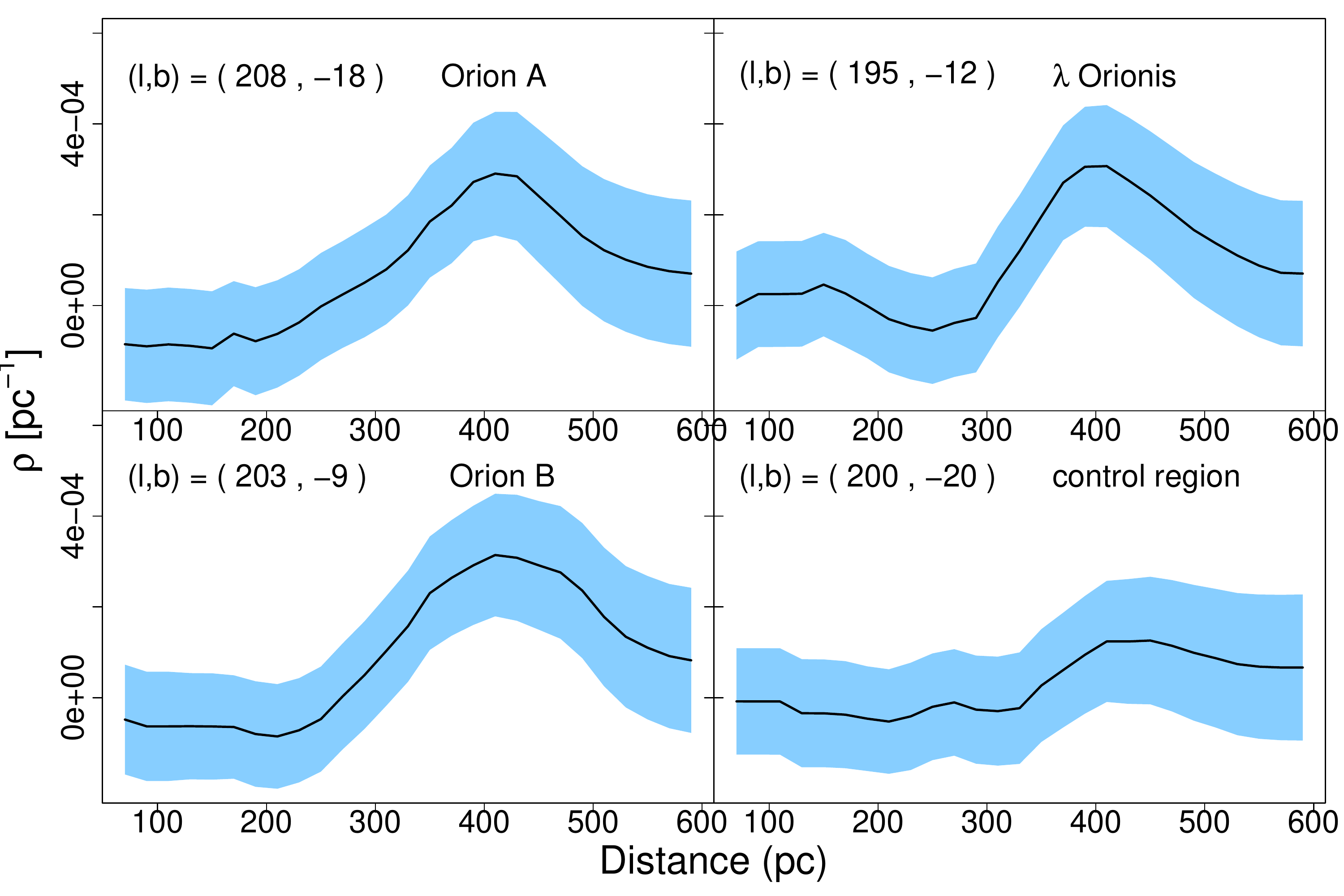}
\caption{Dust density vs. distance for four different l.o.s.: three are towards known regions of Orion and one towards a control region where we don't expect a cloud. The black line shows the mean and the blue shades represent one standard deviation (also computed by the Gaussian Process model).}
\label{fig:los_orion}
\end{center}
\end{figure}

Our model predicts a Gaussian distribution for our knowledge of the dust density at every point in space, 
from which we can extract an estimate (the mean, which has been plotted so far) and its uncertainty (standard deviation).
Figure \ref{fig:los_orion} shows the variation of both of these quantities along different lines-of-sight. Each panel of the figure shows one l.o.s towards Orion A, Orion B, and $\lambda$ Orionis, plus one l.o.s outside these regions where we do not expect to see a prominant peak in the dust density. The dust density in Orion A and B starts increasing slightly after 200 pc, while dust towards $\lambda$ Orionis seems to be more concentrated, increasing only after 300 pc. The dust density in Orion A decreases to the input mean value by a distance of 500 pc; however, in Orion B the density remains high to larger distances which demonstrates a more extended dense part at the direction of Orion B.

\begin{figure} 
\begin{center}
\includegraphics[width=0.50\textwidth, angle=0]{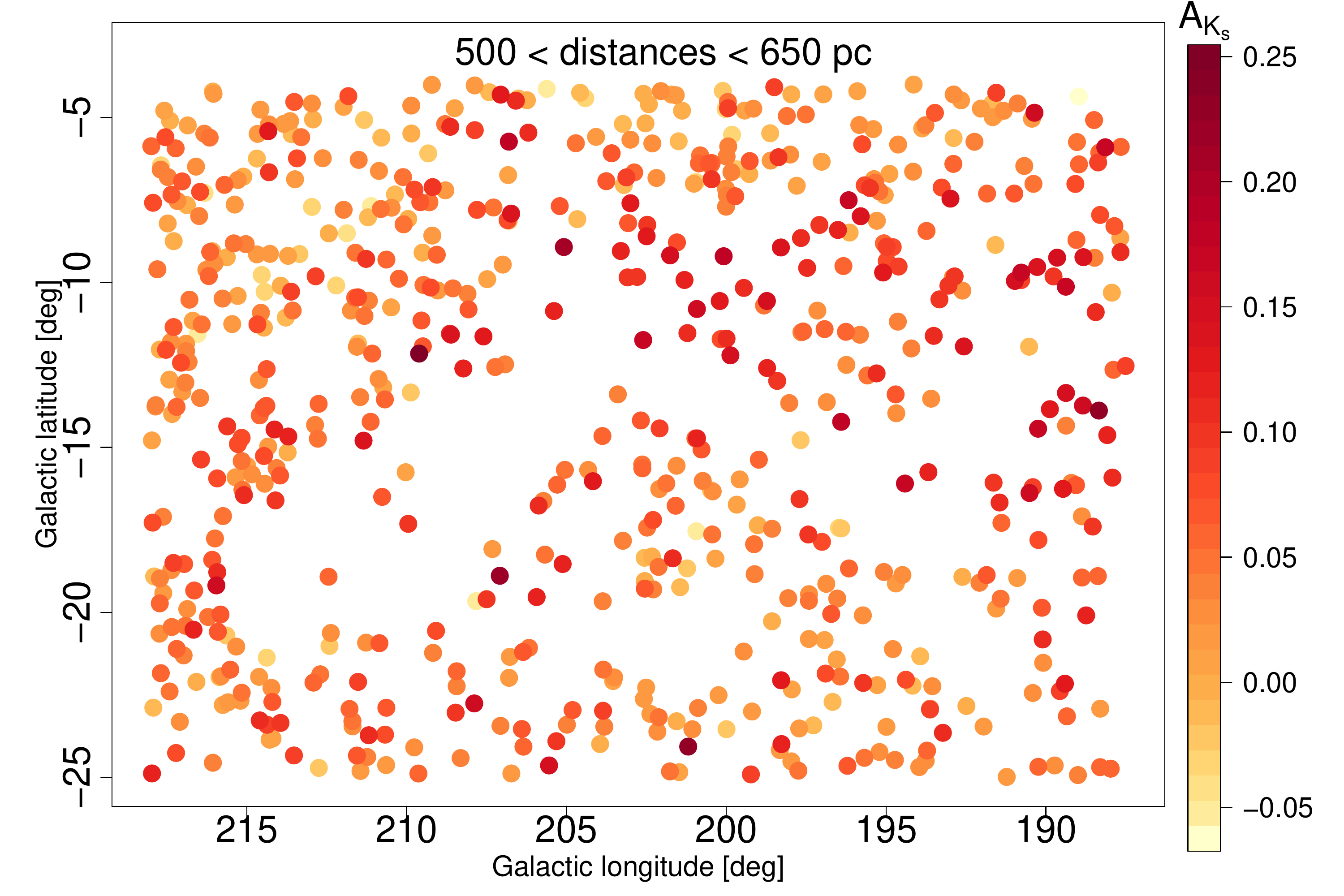}
\caption{Latitude vs.\ longitude of input stars towards Orion complex further than 500 pc colour-coded by their input extinctions. A region extending down and left of centre and down and right of centre is devoid of stars due to the high foreground extinction. The extinctions of the stars at the edge of this region are increasing towards the centre of the region.}
\label{fig:500_orion}
\end{center}
\end{figure}

Figure \ref{fig:500_orion} shows the input stars which lie behind the Orion cloud (further than 500 pc), colour-coded by their extinctions. The first feature from this plot is the gap -- missing stars -- extending from the centre of the plot to the lower left and right. This is a consequence of stars behind the dense regions of the Orion complex being highly extinct and thus too faint for our data sample, on account of the magnitude limit of Gaia-TGAS ($G \simeq 13.5$, although incompleteness sets it at brighter magnitudes). The second feature is the increase of the extinction toward the edges of these missing parts,
from which we can better identify the location of the foreground obscuring cloud. However, this makes our model underestimate the predictions in these highly extinct regions where stars are missing, and be uncertain about the back edges of the clouds.

\section{Discussion}\label{discussion}

\begin{figure*}
\resizebox{\hsize}{!}{\includegraphics[clip=true]{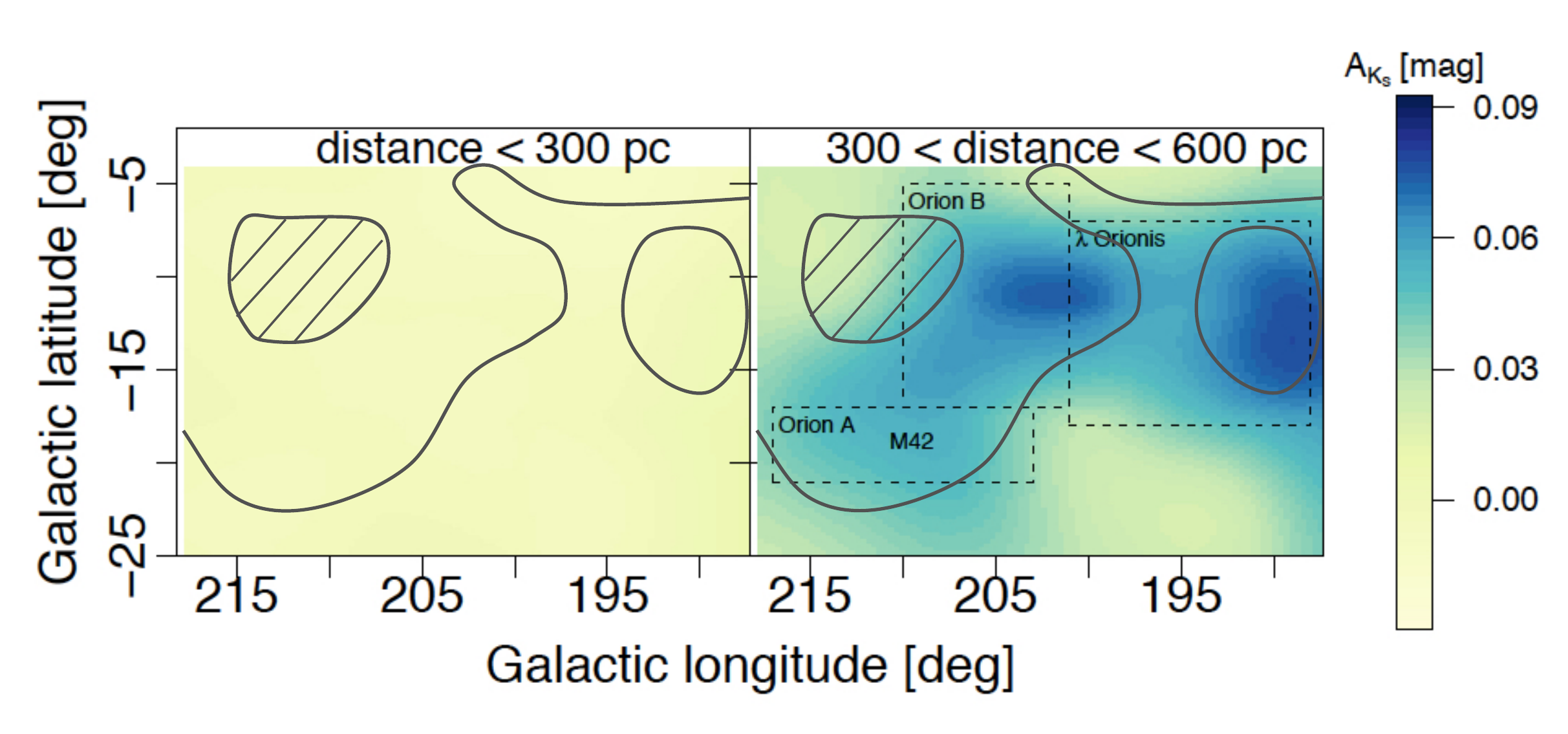}}
\caption{Integrated dust density predictions (i.e.\ extinctions $A_{K_{s}}$ in mag.) for the Orion complex for the same two distance slices as shown in figure 2 of \cite{schlafly15}. Most of the extinction is at distances more than 300 pc. Orion A, Orion B, and $\lambda$ Orionis are marked by dashed lines \citep{Lombardi11} and the solid contours show the E(B-V) = 0.5 mag from the 2D dust map of \citet*{Schlegel98} (SFD) where the hatched area indicates lower reddening.}
\label{fig:300_orion}
\end{figure*}

Our results indicate that the Orion molecular complex includes a higher density dust region starting as close as about 300 pc and extending up to around 500 pc, suggesting a depth of 200 pc for the complex. We compare our results with the recent work of \cite{schlafly15} in figure \ref{fig:300_orion}, which shows the same distance slices as in the first two panels of figure 2 of \cite{schlafly15}. The colour shows the integrated dust densities through the cloud ($A_{K_{s}}$ in mag. which is approximately 0.1 times $A_{V}$). There is a good overall agreement: Most of the dust we infer is at distances between 300 pc and 600 pc and the three main parts related to Orion A, Orion B, and $\lambda$ Orionis are obvious in the right panel. We also over-plot contours of E(B - V) = 0.5 mag from the \citet*{Schlegel98} (SFD) 2D reddening map. This matches our dust density predictions pretty well, except for the regions closer to the Galactic disk and larger longitudes where the dust probed by SFD is further away and beyond our probed distance range.

If we compare these reconstructed extinctions towards the Orion complex with the input extinctions (figure \ref{fig:input_orion}), we see the maximum $A_{K_{s}}$ value in figure \ref{fig:300_orion} is only about 0.1 mag while the input extinctions extend to 0.6 mag. This is the expected outcome of our isotropic Gaussian process prior with a particular scale length (100 pc here) which considers correlations in all directions. This means that each point will be affected by the surrounding points within the 100-pc correlation length. Therefore, in order to compare the predicted extinctions with the input ones, we need to have an estimation of the average extinctions within the 100-pc radius in figure \ref{fig:input_orion}.
The average extinction over 100-pc radius for most of the regions in figure \ref{fig:input_orion} is around $0.05\pm0.05$ mag which is in agreement with the reconstructed extinctions.

It is important to realise that some dust density predictions are negative. This does not have a physical meaning, but it is a consequence of the Gaussian process assumption and the noisy input data. We could put some stronger constraints (priors) on the predictions (e.g.\ with truncated functions) to get only positive values. But we chose not to do this, as these negative values are informative and show inconsistencies in the input extinctions and/or distances. This is because the integral over dust density predictions along the l.o.s towards each star needs to be equal to the input extinction for the likelihood model. Therefore, if there are some extinctions which cause the model to predict higher dust densities than what is expected for more distant stars (along or near to the l.os.), then the model has to predict some negative dust densities to stay consistent with other input extinctions within their error bars. This is evident from figure \ref{fig:front} where not only the impact of the very high input extinction values appears as an increase in the maximum predicted extinction, but also they produce maps with much larger negative dust densities than before, which is another indication that the some of the input extinctions are unrealistically high.

Another point to consider is that by using the Gaia-TGAS catalogue, we underestimate the amount of dust in the Orion complex. This is because even a modest amount of extinction means stars will not appear in Gaia-TGAS, due to its relatively bright magnitude limit ($G \simeq 13.5$). Thus we can only see the cloud to a limited depth. This will improve with the use of Gaia DR2, which will go around seven magnitudes deeper than the TGAS catalogue. With Gaia DR2 we will be able to make a deeper and more precise 3D view of the Orion cloud and infer more accurate distances towards its different parts. Moreover, the extinction estimates in Gaia DR2 have uncertainties of order 0.2 mag in the G band (Andrae et al. 2018 A\&A submitted), which is equivalent to about 0.02 mag in the K-band: this is five times smaller as used in this study, although there are additional issues (such as systematics) with using the Gaia DR2 extinctions.

\subsection{Spurious high-extinction sources}\label{high}
\begin{figure} 
\begin{center}
\includegraphics[width=0.50\textwidth, angle=0]{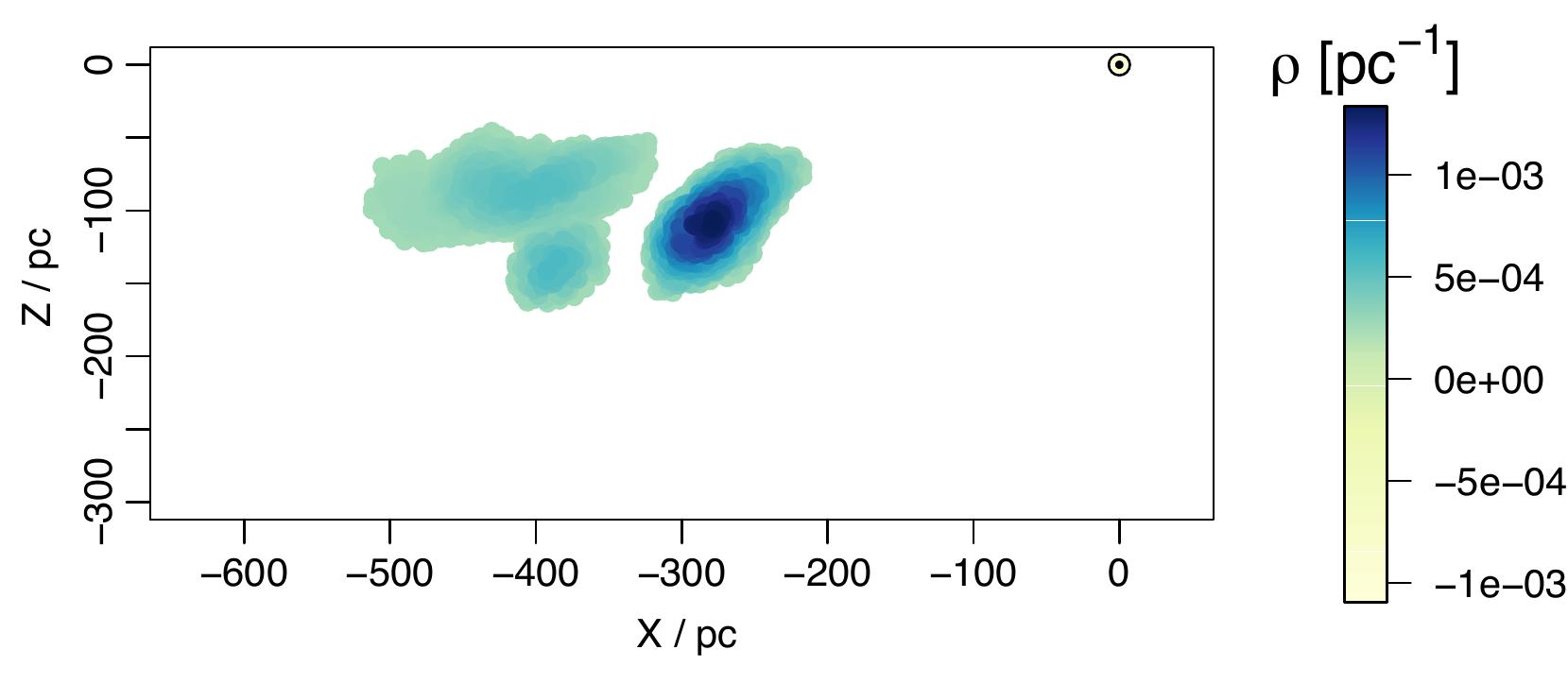}
\caption{The effects of the spuriously blue stars with extremely high extinction values on the predictions; there is a foreground cloud in front of Orion. The axis and view angle are the same as in figure \ref{fig:3D_orion}, right panel.}
\label{fig:front}
\end{center}
\end{figure}

As mentioned briefly in section \ref{data}, using the observed colours of stars and RJCE method results in having extremely high extinction values for 19 out of 13\,000 stars, mostly located at the lower centre of the second and third distance slices of figure \ref{fig:input_orion} which we decided to discard based on the CMD and the theoretical models. To study these stars we include them in the sample and predict the dust densities one more time to see their effects on the results. Figure \ref{fig:front} shows the resulting map. The main change 
(cf.\ \ref{fig:3D_orion}, right panel) is the appearance of a distinct cloud in front of Orion A.
Looking up these stars on SIMBAD suggests that many of these supposedly high-extinction sources are young-stellar objects in the foreground of Orion and are associated with bursts of star formation in Orion over the last several million years \citep[e.g.][]{Bouy14,Bally08,Brown94,Brown95}. These stars have mid-infrared excesses, which is due to their circumstellar dust rather than interstellar dust, leading to spurious extinction predictions from RJCE, for which stars are assumed to be older. Another reason for having such extreme, unlikely extinction values is our approach to deriving colours for these stars. Since we obtain NIR and MIR photometry from different surveys (2MASS and WISE), and as these surveys did not collect their data at the same time, stellar variability could result in erroneous colours ($H - K_{s}$ colour of more than 2 magnitudes). We therefore used the location of stars in the CMD to help assess the validity of their dereddened colours. This experiment demonstrates that using only colours of stars without other constrains, like the CMD, can cause unexpected results, because the colours might have been affected by different factors which could, in our case, noticeably affect the predictions.

\subsection{Negative input extinction effects}\label{negative}
\begin{figure}
\begin{center}
\includegraphics[width=0.50\textwidth, angle=0]{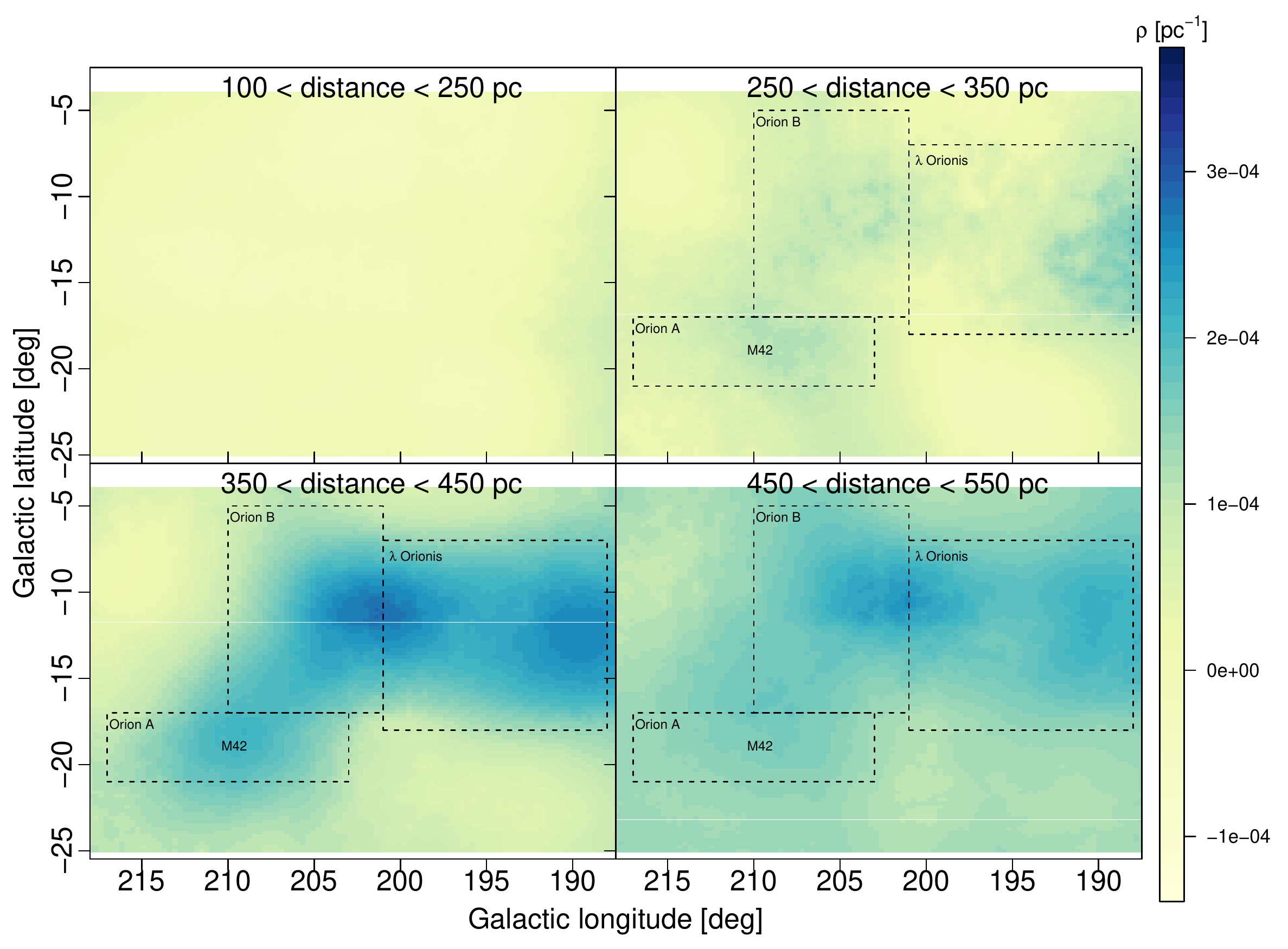}
\caption{Dust density predictions as in figure \ref{fig:pred_2by2_orion} but using only stars with positive extinctions as the input. The colour bar is the same as in figure \ref{fig:pred_2by2_orion} enabling direct comparison. Orion A, Orion B, and $\lambda$ Orionis are marked by dashed lines \cite{Lombardi11}.}
\label{fig:pos}
\end{center}
\end{figure}

\begin{figure} 
\begin{center}
\hspace*{-2em}\includegraphics[width=0.50\textwidth, angle=0]{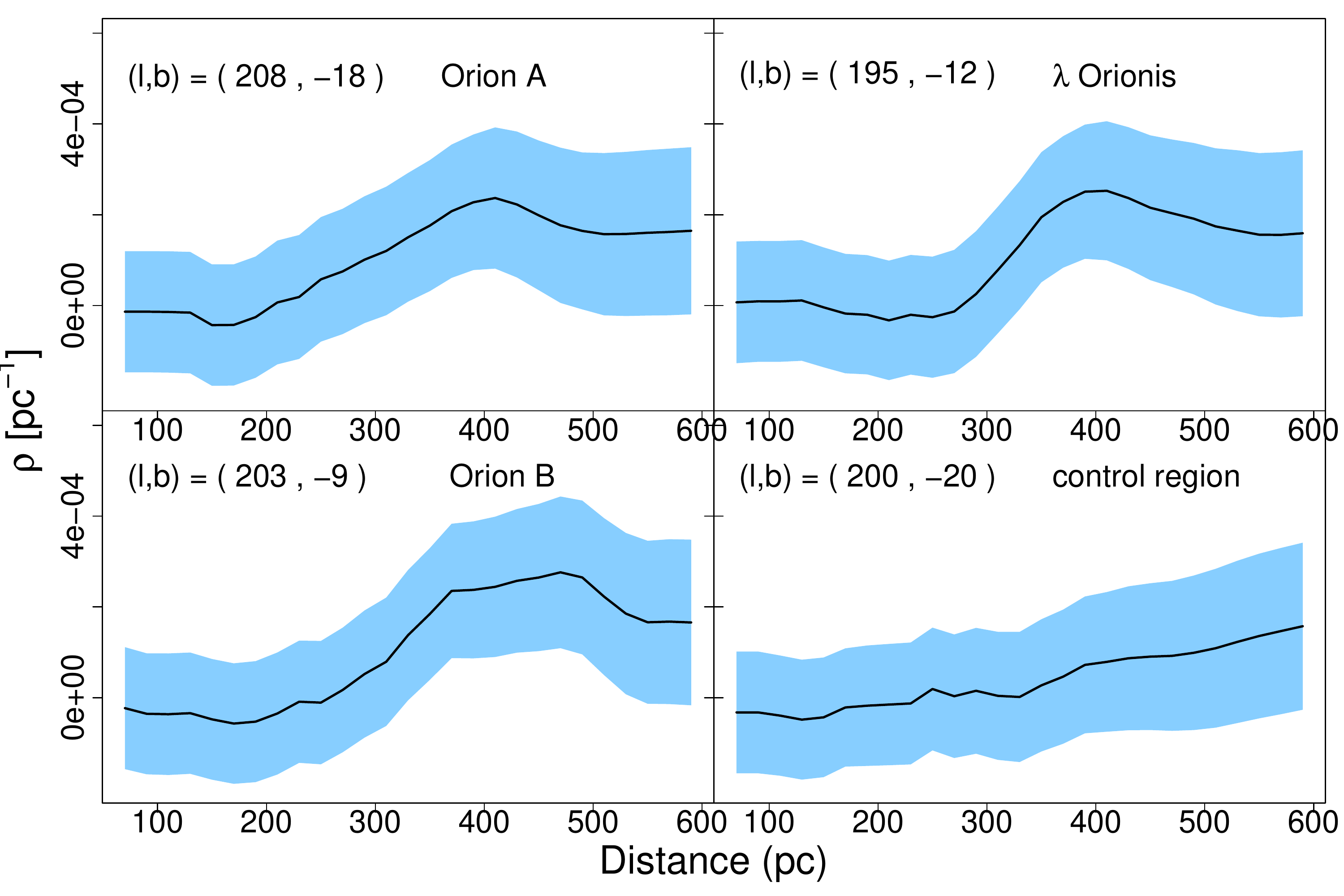}
\caption{A figure \ref{fig:los_orion} but using the positive extinctions to infer the dust densities.}
\label{fig:los_pos}
\end{center}
\end{figure}

As mentioned in section \ref{data}, around $40\%$ of stars in our sample have negative extinctions (see figure \ref{fig:hist}) which we included them in our analysis so far. To see the effects of these negative extinctions, we remove them from the sample and redo the analysis using only positive extinctions. This makes the input mean density almost twice as high as when including negative ones. Figure \ref{fig:pos} shows the predictions using only positive extinctions with the same colour range as in figure \ref{fig:pred_2by2_orion}. The results look rather similar: they trace similar structures in the dust densities, although the range of the dust density is now narrower. This can be seen better by comparing figures \ref{fig:los_orion} and \ref{fig:los_pos}, where figure \ref{fig:los_pos} shows dust density predictions per l.o.s for the case of using only positive extinctions. This decrease in dust amplitude arises because discarding negative extinctions decreases the contrast between high and low density regions. Excluding negative extinctions produces less negative dust densities overall.


\subsection{Input extinction uncertainties}\label{error}
\begin{figure} 
\begin{center}
\hspace*{-2em}\includegraphics[width=0.50\textwidth, angle=0]{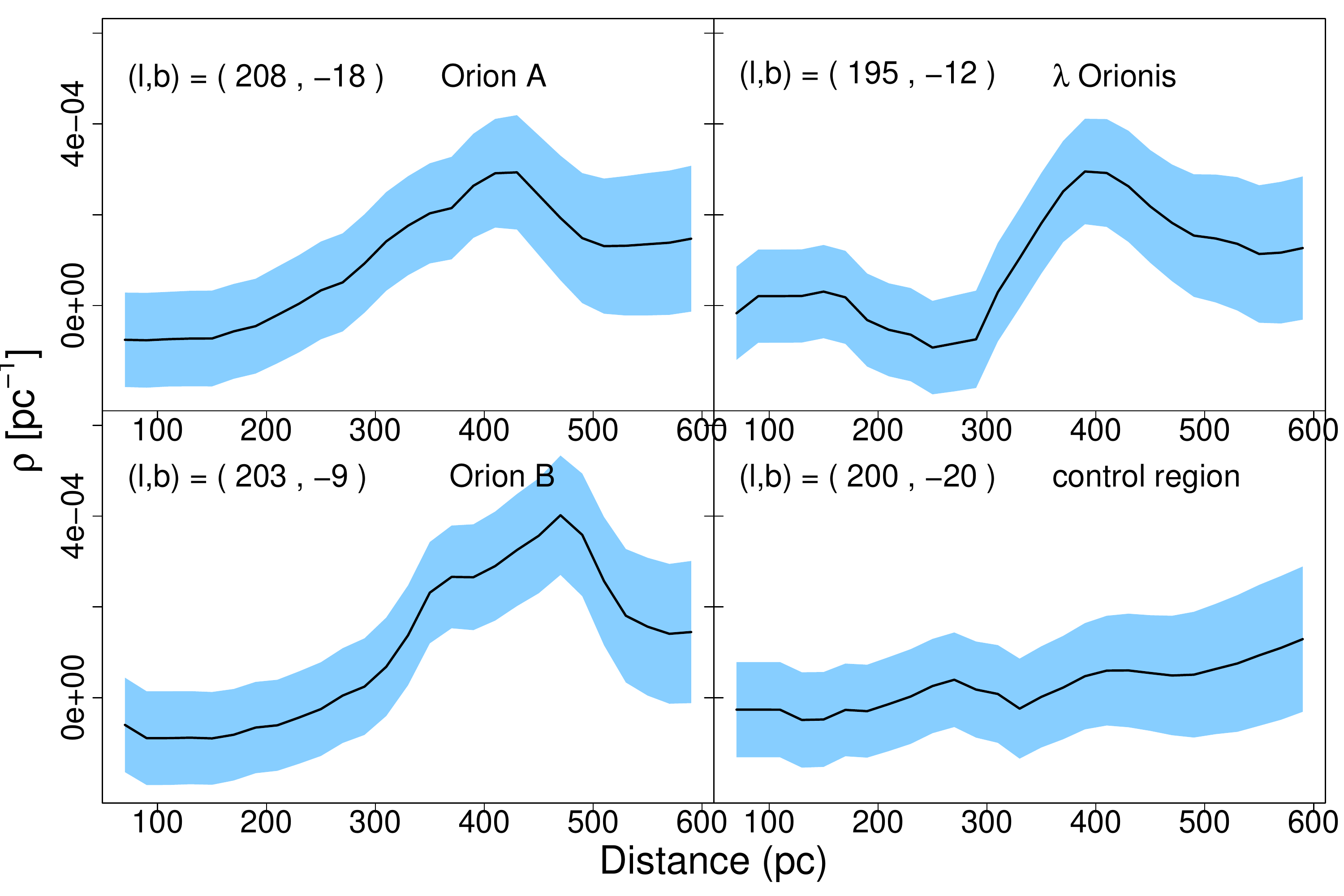}
\caption{As figure \ref{fig:los_pos} but from using uncertainties on the extinctions of 0.05 mag as opposed to 0.10 mag.}
\label{fig:los_pos0.05}
\end{center}
\end{figure}

As mentioned in section \ref{data}, we have assumed a constant extinction uncertainty of 0.1 mag (the upper limit of the uncertainties expected from the RJCE method) for all input stars. Here we investigate the effects of input extinction uncertainties on our density predictions. Figure \ref{fig:los_pos0.05} shows l.o.s predictions as in figure \ref{fig:los_pos} using only positive extinctions with a smaller uncertainty of 0.05 mag. The predictions are less smooth and the error bars are smaller. This is as expected: having smaller uncertainties in the extinction values indicates a sharper likelihood which therefore has more impact on the posterior than does our Gaussian process prior. There are also more negative density predictions at lower distances which are due to the fact that dropping the uncertainties make input extinctions less consistent within their error bars, therefore predicting negative values at lower density regimes.

In conclusion, we have inferred the 3D structure of the Orion complex using astrometry from Gaia-TGAS combined with photometry from 2MASS and WISE. We have estimated the distances and the depth of the cloud to be compatible with other recent works, which demonstrates that this approach can be applied on the local molecular clouds to trace their 3D dust structures. However, the distance to the back of the Orion cloud is likely underestimated due to the restricted depth of our input catalogues. Having more accurate distance derived from Gaia DR2 parallaxes, plus better extinction measurements, we will map the 3D spatial shapes of several local molecular clouds to reveal their possible physical connections in 3D space.

\section*{Acknowledgements}
We would like to thank the anonymous referee for their constructive comments. This work has made use of data from the European Space Agency (ESA)
mission {\it Gaia} (\url{https://www.cosmos.esa.int/gaia}), processed by
the {\it Gaia} Data Processing and Analysis Consortium (DPAC,
\url{https://www.cosmos.esa.int/web/gaia/dpac/consortium}). Funding
for the DPAC has been provided by national institutions, in particular
the institutions participating in the {\it Gaia} Multilateral Agreement.
This project is partially funded by the Sonderforschungsbereich SFB\,881 ``The Milky Way System'' of the German Research Foundation (DFG). 

\bibliographystyle{aa}
\bibliography{Rezaei_Kh._2017_2}

\end{document}